\documentclass[prl,aps,twocolumn]{revtex4}
\usepackage{amsmath,graphics,epsfig,color,verbatim}

\begin{document}

\title{Na induced correlations in Na$_x$CoO$_2$}

\author{C. A. Marianetti$^1$ and G. Kotliar$^{1,2,3}$ }

\date{\today}

\begin{abstract}
Increasing experimental evidence is building which indicates that
signatures of strong correlations are present in the Na rich region
of Na$_x$CoO$_2$ (ie. $x\approx0.7$) and absent in the Na poor region (ie. $x\approx0.3$). 
This is unexpected given
that NaCoO$_2$ is a band insulator and CoO$_2$ has an integer filled open shell  making
it a candidate for strong correlations.
We explain these experimental observations by  presenting  a minimal low-energy 
Hamiltonian for the cobaltates and solving it within LDA+DMFT.
The Na potential is 
shown to be a key element in understanding correlations in this material.
Furthermore, LDA calculations for the realistic Na ordering predict a \emph{binary} perturbation of the Co sites 
which correlates with the Na$_1$ sites (ie. Na sites above/below Co sites).
\end{abstract}

\address{$^1$Department of Physics and Astronomy and Center for Condensed Matter
Theory, Rutgers University, Piscataway, NJ 08854--8019}
\address{$^2$ Service de Physique Theorique, CEA Saclay, 91191 Gif-Sur-Yvette, France}
\address{$^3$ Centre de Physique Theorique, Ecole Polytechnique 91128 Palaiseau Cedex, France}

\maketitle

The qualitative features of the cobaltate
phase diagram have been experimentally established \cite{Foo:2004}. Uncorrelated behavior is observed in the
Na poor region while correlated behavior is observed in the Na rich
region. For example, Magnetic susceptibility measurements show Pauli-like behavior for 
$x\approx0.3$ and Curie-Weiss behavior for 
$x\approx0.7$
\cite{Foo:2004} 
(for further references see \cite{Ihara:2004}).

These observations are counterintuitive given our conventional 
understanding of Na$_x$CoO$_2$. In NaCoO$_2$, Co is in the 3+ configuration and thus has
six electrons. The cubic component of the crystal field splits the $d$ manifold into a set
of 3-fold $t_{2g}$ orbitals and 2-fold $e_g$ orbitals, while the trigonal component 
will further split the $t_{2g}$ orbitals into $a_{1g}$ and $e_g'$. In this scenario, the six
electrons of the Co will fill the $t_{2g}$ orbitals (ie. $a_{1g}+e_g'$) resulting in a band 
insulator. 
On the contrary,
in CoO$_2$ the Co will be in a 4+ configuration having 5 electrons in the $t_{2g}$ shell, and
may either be a metal or a Mott insulator. 

There are two main puzzles posed by the above listed experimental observations.
First, given that NaCoO$_2$ is a band insulator, it is difficult to understand why correlations are observed
for the nearby composition of $x=0.7$. The density of holes will be relatively low 
and therefore the on-site coulomb repulsion will have a minimal effect.
Second, one would expect correlations to increase as the system is doped
towards $x=0.3$ as the hole density in the $t_{2g}$ shell is increasing towards integer occupancy with an open shell. 
We resolve this puzzle by 
proposing a minimal low energy Hamiltonian which captures the essential physics and
solve it within LDA+DMFT.

\begin{eqnarray}\label{ham}
H=\sum_{ij\alpha\beta\sigma}t_{\alpha\beta}c^+_{i\alpha\sigma}c_{j\beta\sigma}+
\sum_{i\alpha\beta\sigma\sigma'}U_{\alpha\beta}^{\sigma \sigma'}n_{i\alpha\sigma}n_{i\beta\sigma'}
%\nonumber\\
+\sum_{i\alpha\sigma}\epsilon_{\alpha i} n_{\alpha i}
\end{eqnarray}

where $i,j$ are site indices, $\alpha,\beta$ are orbital indices running over $a_{1g}$ and
$e_g'$, $t$ is the hopping parameter, $U$ is the on-site Coulomb repulsions, and $\epsilon$ 
is the on-site potential which mimics the Na potential. It should be emphasized that the orbitals to which 
the creation operators refer are low energy orbitals, effectively
composed of oxygen $p$ orbitals as well as Co $d$ orbitals which compose the $t_{2g}$ anti-bonding manifold. 
All the parameters of the Hamiltonian are in general 
functions of the total density, as the system rehybridizes \cite{Marianetti:2004} as a function of doping.

\begin{figure}[h]
\includegraphics[angle=-90, width=\linewidth, clip=, bb= 42 35 428 322]{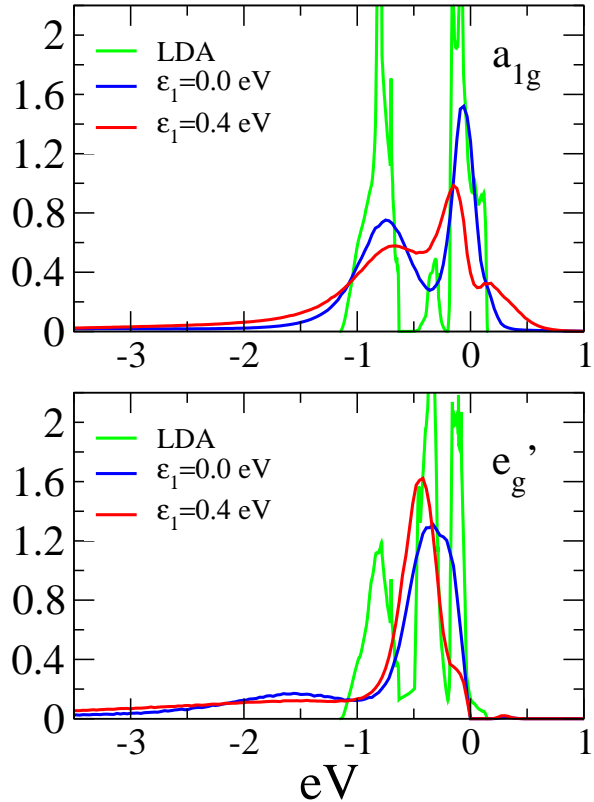}
\caption{Spectral functions for Na$_{0.7}$CoO$_2$ with random binary disorder for the $a_{1g}$ and $e_g'$ 
orbitals for T=290 K.
} \label{fig:spectraldisorder}
\end{figure}

\begin{figure}[h]
\includegraphics[angle=-90, width=\linewidth, clip=, bb= 42 35 428 322]{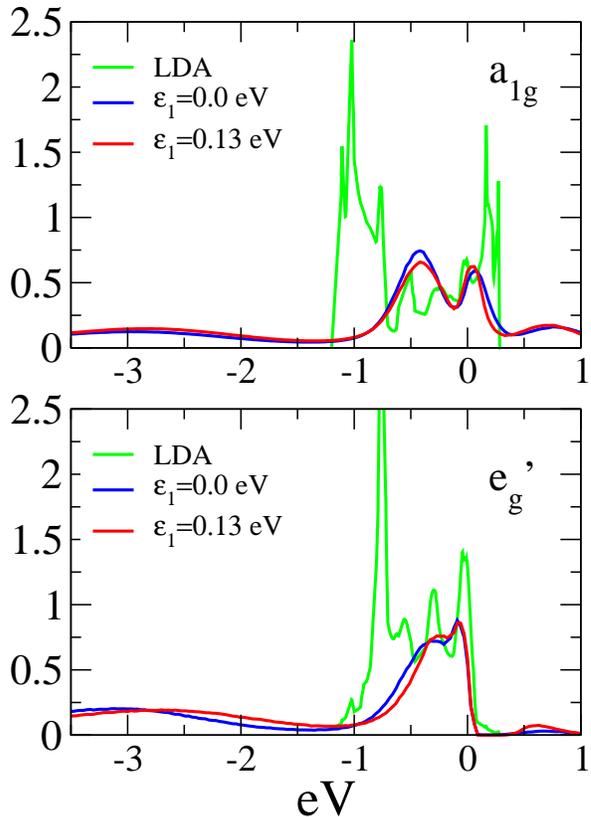}
\caption{Spectral functions at Na$_{0.3}$CoO$_2$ for the $a_{1g}$ and $e_g'$ orbitals for T=290 K. 
} \label{fig:spec_x0.3}
\end{figure}

The hopping elements of the Hamiltonian are obtained using 
LDA calculations on NaCoO$_2$  and CoO$_2$ to be used for DMFT calculations in the Na rich region and Na poor region, 
respectively. The projected DOS for the 
$a_{1g}$ and $e_g'$ states are shown in figure \ref{fig:spectraldisorder} and \ref{fig:spec_x0.3}, respectively. 
In both cases, the $a_{1g}$ orbital
has the overwhelming fraction of hole density. 
The doping dependence of the hopping parameters is substantial given the changes
in the shape of the DOS and given that the total bandwidth increases by 0.25 $eV$ for CoO$_2$.
Detailed cluster calculations for LiCoO$_2$ fit the on-site $U$ 
to the experimental photoemission yielding $U=3.5 eV$ \cite{Vanelp:1991}. This should be very similar
to NaCoO$_2$, and we use a slightly 
reduced value of  $U=3 eV$ because
we are considering low-energy orbitals.
Additionally, we assume $U$ to be
independent of doping.
A nearest neighbor Coulomb interaction should be added to eq \ref{ham}
in order to assist in forming charge ordered phases \cite{Motrunich:2004B},
and in single-site DMFT this results in a Hartree shift.

The on-site potential $\epsilon$ is treated as a \emph{random binary} variable, 
and the fact that it is binary will justified in the  LDA calculations  below.
Treating the potential as random should be accurate for average quantities in the Na ordered phases, 
or even k-resolved quantities in the disordered region of the phase diagram.  
A fraction $x_1$ of the lattice sites will have a on-site potential $\epsilon_1$ while
$(1-x_1)$ sites will have a have a on-site potential $\epsilon_2=\frac{x_1\epsilon_1}{x_1-1}$.
We will assume 
that $x_{\epsilon_1}$ corresponds to $1-x_{Na}$, where $x_{Na}$ is the fraction of 
Na in the material. This also will be justified by our LDA calculations.

We propose that the Na potential plays a critical role in the behavior of this material.
In a previous study of Li$_x$CoO$_2$ \cite{Marianetti:2004}(a very similar material), LDA calculations demonstrated that
in the dilute Li-vacancy region of the phase diagram, the Li vacancy binds the hole which is
usually doped into the t$_{2g}$ band and hence forms a half-filled impurity band split 
off from the valence band (analogous to
Phosphorous doped Silicon, except the vacancies are highly mobile). Due to the fact 
that the impurity band is half-filled, it may be strongly correlated
and form a Mott insulator if the on-site interaction is sufficiently strong.
This behavior persists until enough holes are present
to screen the Li vacancies upon which the more standard picture of doping into the $t_{2g}$
states is recovered. The LDA calculations predict the impurity band to completely merge into
the valence band once $x\approx0.98$. Due to the similarities of the two materials, it is likely that
a similar scenario occurs in Na$_x$CoO$_2$. Of course, the experiments of interest are in a
region of much higher doping (ie. $x\approx0.7$) where it is implausible that an impurity band would persist.
However, it is likely that the Na vacancies will still act as a strong
perturbation towards the Co sites. Therefore, it is possible that the Na vacancy potential
will cause certain Co sites to be higher in energy and possess a higher occupation
than the average number of holes. 
Increasing the occupation of a given site towards integer filling will
favor moment formation assuming that the on-site coulomb repulsion is sufficiently strong.
This phenomena has been studied in detail in the context of model Hubbard 
Hamiltonians \cite{Byczuk:2003}. Merino et al.
suggested a Hubbard model on a triangular lattice in the presence of an ordered potential
in order to explain the correlated behavior observed in Na$_x$CoO$_2$ \cite{Merino:2005}.
They note that in the limit of strong coupling of the ordered potential that 
correlations will be formed, similar to our work.
NMR measurements strongly support this notion of certain Co sites being favored as
there are distinct Co sites for $x=0.7$ while there are not for $x=0.3$
\cite{Mukhamedshin:2005}. 
The fact that only one site is observed at $x=0.3$ supports the notion that the Na potential is well screened
an no longer a significant perturbation for small $x$.

\begin{figure}[htb]
\includegraphics[width=\linewidth,clip= ]{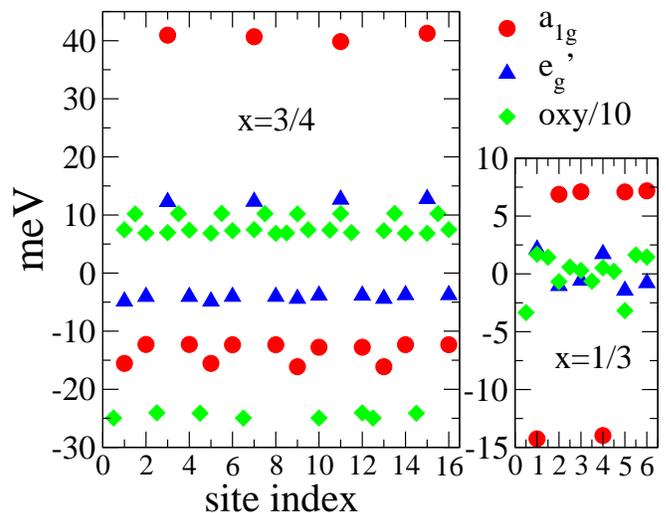}
\caption{The on-site orbital energy,$m_{i \alpha}$, relative to the 
average for various orbitals at each site in the unit cell.
The energies for the oxygen orbitals have been divided by ten. The left and right panels
correspond to Na$_{\frac{3}{4}}$CoO$_2$ and Na$_{\frac{1}{3}}$CoO$_2$, respectively.
} \label{fig:siteenerg}
\end{figure}

In order to understand the effect of the Na 
potential described by the
on-site potential $\epsilon$
in eq. \ref{ham},  we performed
LDA calculations on the $x=\frac{1}{3}$ and $x=\frac{3}{4}$ unit cells which were experimentally determined by
Zandbergen et al\cite{Zandbergen:2004}. These unit cells have 6 and 16 formula  units for
the $x=\frac{1}{3}$ and $x=\frac{3}{4}$ unit cells, respectively, and
both contain equal numbers of occupied Na$_1$ and Na$_2$ sites. 
The Na$_1$ sites project
directly onto the triangular Co lattice while the  Na$_2$ sites project onto the centroids
of the Co triangular lattice \cite{Zandbergen:2004}.
 All LDA calculations were performed using the Vienna Ab-Initio Simulation Code (VASP) \cite{Kresse:1996}.
In order to quantify the effect of the Na potential on   the different orbitals, 
the first moment $m_{i \alpha}$ of the projected DOS (ie. the on-site orbital energy)
of the Oxygen and Cobalt  orbitals are calculated for each site of the supercell.
$\label{moment}
m_{i \alpha}=\frac{\int \rho_{i \alpha}(\epsilon)\epsilon d \epsilon}{\int \rho_{i \alpha}(\epsilon) d \epsilon}
$
where the indices $i$ and $\alpha$ are the site and  orbital indices,  respectively. The deviation of the orbital
energy from the site average is shown in figure \ref{fig:siteenerg} (ie $m_{i \alpha}-\bar m_{\alpha}$).
Remarkably, the results are  roughly  binary distributions as evidenced by the horizontal lines which are 
formed by the data points. The Na ordering causes each site to be perturbed, resulting in either a high or low
energy orbital on a given site. The oxygen orbitals are perturbed the most followed by the a$_{1g}$ and e$_g'$ orbitals.
The perturbations are substantially smaller for x=$\frac{1}{3}$ as compared to x=$\frac{3}{4}$, which is
expected given that a larger amount of hole density is present at  x=$\frac{1}{3}$ to screen the Na potential.
A very simple rule governs the observed behavior. In the x=$\frac{3}{4}$ structure, all of the low energy 
a$_{1g}$ and e$_g'$ orbitals have one nearest-neighbor Na$_1$ present, while the high energy orbitals do not.
There are 6 Na$_1$ sites occupied in this unit cell, each of which will have two Co nearest neighbors, and hence 
there are 4 high energy Co sites and 12 low energy sites.
For the x=$\frac{1}{3}$ structure, the a$_{1g}$ orbital follow the same rule. However, the e$_g'$ orbitals are 
perturbed in the opposite direction in this case. Although, the perturbations of the e$_g'$ orbitals are all
less than $2 meV$ and should be considered carefully.
The splitting of the oxygen orbitals also follow a simple rule in the x=$\frac{3}{4}$ structure.
All of the low energy oxygens have three nearest-neighbor Na while the high energy oxygens only have two
nearest-neighbor Na. For the x=$\frac{1}{3}$ structure, the oxygen do not split into a binary distribution,
but it can be roughly understood as a ternary distribution if one analyzes both nearest and next-nearest neighbor Na.
The low energy oxygen have three neighboring Na, the high energy oxygens have two Na$_1$ neighbors, while the intermediate
energy oxygens have two Na$_2$ neighbors. Regardless, the $a_{1g}$ and $e_g'$ orbitals seem to depend only on the positions
of the Na$_1$.
This seems very reasonable for the a$_{1g}$ orbitals 
given that they point directly towards 
the Na$_1$ sites (ie. $a_{1g}=d_{z^2}$ in the hexagonal coordinate system).
This can be very important for several aspects of the low-energy behavior in this material. For example, this effect
would tend to drive charge ordering to occur on the Co sites which are located directly above/below the occupied Na$_1$ sites.
Additionally, this could effect the stabilization the pockets in the Fermi 
surface as the two orbitals are perturbed
differently. For a recent study on the effect of Na on the Fermi surface see \cite{Singh:2006}.

Obtaining parameters for a model Hamiltonian from LDA calculations is still an open problem. Therefore, we
simply extract the qualitative effects of the Na potential. The binary splitting 
of the oxygen for $x=\frac{3}{4}$ is roughly 350 $meV$ while for $x=\frac{1}{3}$ the maximum splitting is 
roughly 50 $meV$. This should be relevant to the low-energy Hamiltonian given that 
the hole density on the $t_{2g}$ states has associated hole density on the oxygen via the rehybridization mechanism
\cite{Marianetti:2004}. The changes on the Co are smaller, with the $a_{1g}$ splitting being roughly 55 and 20 $meV$
for $x=\frac{3}{4}$ and $x=\frac{1}{3}$, respectively. This is to be expected given that these states are near the Fermi 
energy and are not only well screened, but overscreened by LDA, and thus their bare values may be significantly larger. We take a value of $\epsilon_1=400meV$ for $x=\frac{3}{4}$ and $\epsilon_1=\frac{400}{3}meV$ for $x=\frac{1}{3}$, reflecting
the fact that $x=\frac{1}{3}$ more screened. A more appropriate value for $x=\frac{1}{3}$ 
might be $\epsilon_1=0$ given that only a single Co is observed in NMR \cite{Mukhamedshin:2005}, and we explore this as well. We conservatively assume $\epsilon$ to be orbitally independent, which
will make it more difficult to form correlations near the band insulator. Below we will
demonstrate that these estimates yield predictions consistent with experimental observations.

With these order-of-magnitude estimates, we shall now proceed to solve the proposed Hamiltonian (equation \ref{ham}) 
within LDA+DMFT \cite{Kotliar:2006},
including the additional effect of binary disorder 
(see \cite{Georges:1996} for references). In this case there will be two 
impurity models which represent the two different binary environments.
We begin with a guess for the average hybridization function $\Delta$, and then construct the
bath function $G_0=(i\omega_n - E_{imp} +\mu - \Delta)^{-1}$ where $E_{imp}$ is the average impurity level
and $\mu$ is the chemical potential. We then construct the two disorder
bath functions $G_0^{\epsilon_1}=(G_0^{-1}-\epsilon_1)^{-1}$ and $G_0^{\epsilon_2}=(G_0^{-1}-\epsilon_2)^{-1}$ .
The two corresponding impurity problems are then solved and the average Green's function is 
constructed as $G=x_{\epsilon_1}G_{\epsilon_1}+(1-x_{\epsilon_1})G_{\epsilon_2}$. The
average self-energy is then constructed using Dyson's equation 
$\Sigma(i\omega_n)=G_0^{-1}(i\omega_n)-G^{-1}(i\omega_n)$. Finally,
the DMFT self-consistency condition is performed:

\begin{eqnarray}\label{selfcons}
G_0(i\omega_n)&=\left[ \Sigma(i\omega_n) + 
\left( \sum_k \frac{1}{i\omega_n - H_k + \mu - \Sigma(i\omega_n)} \right)^{-1} \right]^{-1}
\end{eqnarray}

where $H_k$ is the homogeneous Hamiltonian (ie. eq. \ref{ham} with $\epsilon =0$) in k-space. 
The entire process is then repeated until convergence is achieved. Given that the local Green's function
is diagonal (ie. $a_{1g}+e_g'$), the summation over $k$ in eqn. \ref{selfcons} can be replaced
by a Hilbert transform (see \cite{Georges:1996}).

\begin{figure}[h]
\includegraphics[width=2.9 in,angle=-90,clip= ]{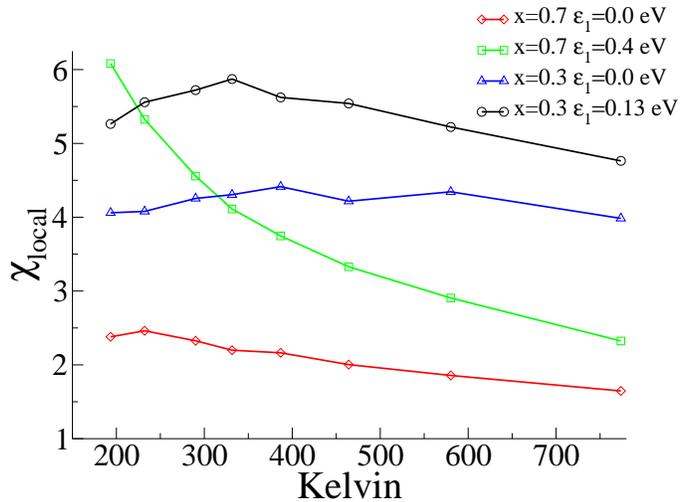}
\caption{Local Magnetic susceptibility versus temperature for Na$_x$CoO$_2$. 
} \label{fig:chi}
\end{figure}

The DMFT impurity problem is solved using Hirsch-Fye Quantum Monte-Carlo (see \cite{Georges:1996,Kotliar:2006})
using the LISA code \cite{Parcollet:2006}. 
All runs were performed with $\frac{\beta U}{L}=1$ and 500,000 Monte-Carlo sweeps.
We begin by considering the Na rich case of $x=0.7$. 
The spectral
functions for the $a_{1g}$ orbital clearly shows the correlations increasing as 
the on-site potential is increased (Figure \ref{fig:spectraldisorder}). 
The states at the Fermi energy are suppressed 
and spectral weight is transferred from low energies to higher energies as the on-site
potential is increased. In both cases, the $e_g'$ orbitals are pushed beneath the Fermi
energy and the pockets are destroyed. 
Another useful quantity to analyze is the local 
magnetic susceptibility (see figure \ref{fig:chi}), which is one of the key experimental measurements. 
The susceptibility is relatively flat for the case with zero on-site potential,
while a Curie-Weiss tail is clearly formed for the case when a on-site potential of 0.4 $eV$ is included.
In that latter case, 30\% of the sites have 0.73 holes while 70\% of the sites
have 0.1 holes. This is qualitatively similar to what is observed
in NMR experiments \cite{Mukhamedshin:2005}. Thus the $\epsilon_1$ site becomes the favored site
receiving significantly more hole density and being driven towards integer-filling where a local moment forms.

Now we proceed to the Na poor region and perform calculations at $x=\frac{1}{3}$. 
The spectral function for the $a_{1g}$ orbital
shows some indication of correlations as some spectral weight has been transferred to higher energies 
(see figure \ref{fig:spec_x0.3}). Additionally, the $e_g'$ orbitals show a small occupation in this case
and the pockets survive as previously shown by Ishida et al \cite{Ishida:2005}. 
The on-site potential only has a small effect on the spectral function.
The main result is that the magnetic susceptibilities only shows a weak temperature dependence in both cases and there 
is no sign of a Curie-Weiss tail (see figure \ref{fig:chi}). 
The on-site potential clearly enhances the susceptibility but is
not capable of building strong correlations. The on-site potential 
creates two distinct sites with 70\% having 0.76 holes and
30\% having 0.5 holes. Both susceptibilities cross the Curie tail of the $x=0.7$ case in the 
vicinity of room temperature, similar to the crossover observed in experiment between 150-250 $K$ \cite{Foo:2004}. 

An additional LDA+DMFT calculation for pure CoO$_2$ 
demonstrates that it is a strongly correlated metal with well formed Hubbard bands and a quasiparticle resonance at the
Fermi energy. 
This guarantees that the susceptibility calculated for Na$_{0.3}$CoO$_2$ will not develop 
a Curie-Weiss tail at low 
temperatures which could not be reached in our QMC calculations.

In conclusion, we have proposed a low-energy Hamiltonian that is capable of explaining the 
observation of correlations near a band insulator and lack of correlation in the Na poor 
region of the doped cobaltates. Realistic parameters for the Hamiltonian obtained from LDA calculations reproduce
the generic behavior observed in the experimental magnetic susceptibility measurements. The Na potential
is shown to be a key element in forming correlations near the band insulator.
LDA calculations for the experimentally predicted Na orderings imply that the Na$_1$ sites 
act as a binary perturbing potential of the Co sites.
Our model 
may be further improved by treating spatial correlation 
of the on-site potential, improved
downfolding techniques to obtain the parameters of the low-energy Hamiltonian, inclusion of  polarons, and the inclusion 
of the $e_g$ states. The $e_g$ states are needed to describe the high energy behavior of
this system, and additionally may cause further renormalization at low energies. Including these
effects may allow for significantly smaller values of the on-site potential $\epsilon$ in order
to achieve the same qualitative effects.

\bibliography{references}

\begin{thebibliography}{15}
\expandafter\ifx\csname natexlab\endcsname\relax\def\natexlab#1{#1}\fi
\expandafter\ifx\csname bibnamefont\endcsname\relax
  \def\bibnamefont#1{#1}\fi
\expandafter\ifx\csname bibfnamefont\endcsname\relax
  \def\bibfnamefont#1{#1}\fi
\expandafter\ifx\csname citenamefont\endcsname\relax
  \def\citenamefont#1{#1}\fi
\expandafter\ifx\csname url\endcsname\relax
  \def\url#1{\texttt{#1}}\fi
\expandafter\ifx\csname urlprefix\endcsname\relax\def\urlprefix{URL }\fi
\providecommand{\bibinfo}[2]{#2}
\providecommand{\eprint}[2][]{\url{#2}}

\bibitem[{\citenamefont{Foo et~al.}(2004)\citenamefont{Foo, Wang, Watauchi,
  Zandbergen, He, Cava, and Ong}}]{Foo:2004}
\bibinfo{author}{\bibfnamefont{M.~L.} \bibnamefont{Foo}},
  \bibinfo{author}{\bibfnamefont{Y.~Y.} \bibnamefont{Wang}},
  \bibinfo{author}{\bibfnamefont{S.}~\bibnamefont{Watauchi}},
  \bibinfo{author}{\bibfnamefont{H.~W.} \bibnamefont{Zandbergen}},
  \bibinfo{author}{\bibfnamefont{T.}~\bibnamefont{He}},
  \bibinfo{author}{\bibfnamefont{R.~J.} \bibnamefont{Cava}}, \bibnamefont{and}
  \bibinfo{author}{\bibfnamefont{N.~P.} \bibnamefont{Ong}},
  \bibinfo{journal}{Phys. Rev. Lett.} \textbf{\bibinfo{volume}{92}},
  \bibinfo{pages}{247001} (\bibinfo{year}{2004}).

\bibitem[{\citenamefont{Ihara et~al.}(2004)\citenamefont{Ihara, Ishida,
  Michioka, Kato, Yoshimura, Sakurai, and Takayama-muromachi}}]{Ihara:2004}
\bibinfo{author}{\bibfnamefont{Y.}~\bibnamefont{Ihara}},
  \bibinfo{author}{\bibfnamefont{K.}~\bibnamefont{Ishida}},
  \bibinfo{author}{\bibfnamefont{C.}~\bibnamefont{Michioka}},
  \bibinfo{author}{\bibfnamefont{M.}~\bibnamefont{Kato}},
  \bibinfo{author}{\bibfnamefont{K.}~\bibnamefont{Yoshimura}},
  \bibinfo{author}{\bibfnamefont{H.}~\bibnamefont{Sakurai}}, \bibnamefont{and}
  \bibinfo{author}{\bibfnamefont{E.}~\bibnamefont{Takayama-muromachi}},
  \bibinfo{journal}{J. Phys. Soc. Jpn.} \textbf{\bibinfo{volume}{73}},
  \bibinfo{pages}{2963} (\bibinfo{year}{2004}).

\bibitem[{\citenamefont{Marianetti et~al.}(2004)\citenamefont{Marianetti,
  Kotliar, and Ceder}}]{Marianetti:2004}
\bibinfo{author}{\bibfnamefont{C.~A.} \bibnamefont{Marianetti}},
  \bibinfo{author}{\bibfnamefont{G.}~\bibnamefont{Kotliar}}, \bibnamefont{and}
  \bibinfo{author}{\bibfnamefont{G.}~\bibnamefont{Ceder}},
  \bibinfo{journal}{Nature Materials} \textbf{\bibinfo{volume}{3}},
  \bibinfo{pages}{627} (\bibinfo{year}{2004}).

\bibitem[{\citenamefont{Vanelp et~al.}(1991)\citenamefont{Vanelp, Wieland,
  Eskes, Kuiper, Sawatzky, Degroot, and Turner}}]{Vanelp:1991}
\bibinfo{author}{\bibfnamefont{J.}~\bibnamefont{Vanelp}},
  \bibinfo{author}{\bibfnamefont{J.~L.} \bibnamefont{Wieland}},
  \bibinfo{author}{\bibfnamefont{H.}~\bibnamefont{Eskes}},
  \bibinfo{author}{\bibfnamefont{P.}~\bibnamefont{Kuiper}},
  \bibinfo{author}{\bibfnamefont{G.~A.} \bibnamefont{Sawatzky}},
  \bibinfo{author}{\bibfnamefont{F.~M.~F.} \bibnamefont{Degroot}},
  \bibnamefont{and} \bibinfo{author}{\bibfnamefont{T.~S.}
  \bibnamefont{Turner}}, \bibinfo{journal}{Phys. Rev. B}
  \textbf{\bibinfo{volume}{44}}, \bibinfo{pages}{6090} (\bibinfo{year}{1991}).

\bibitem[{\citenamefont{Motrunich and Lee}(2004)}]{Motrunich:2004B}
\bibinfo{author}{\bibfnamefont{O.~I.} \bibnamefont{Motrunich}}
  \bibnamefont{and} \bibinfo{author}{\bibfnamefont{P.~A.} \bibnamefont{Lee}},
  \bibinfo{journal}{Phys. Rev. B} \textbf{\bibinfo{volume}{69}},
  \bibinfo{pages}{214516} (\bibinfo{year}{2004}).

\bibitem[{\citenamefont{Byczuk et~al.}(2003)\citenamefont{Byczuk, Ulmke, and
  Vollhardt}}]{Byczuk:2003}
\bibinfo{author}{\bibfnamefont{K.}~\bibnamefont{Byczuk}},
  \bibinfo{author}{\bibfnamefont{M.}~\bibnamefont{Ulmke}}, \bibnamefont{and}
  \bibinfo{author}{\bibfnamefont{D.}~\bibnamefont{Vollhardt}},
  \bibinfo{journal}{Phys. Rev. Lett.} \textbf{\bibinfo{volume}{90}},
  \bibinfo{pages}{196403} (\bibinfo{year}{2003}).

\bibitem[{\citenamefont{Merino et~al.}(2005)\citenamefont{Merino, Powell, and
  Mckenzie}}]{Merino:2005}
\bibinfo{author}{\bibfnamefont{J.~.} \bibnamefont{Merino}},
  \bibinfo{author}{\bibfnamefont{B.~. J.~.} \bibnamefont{Powell}},
  \bibnamefont{and} \bibinfo{author}{\bibfnamefont{R.~. H.~.}
  \bibnamefont{Mckenzie}}, \bibinfo{journal}{Cond-mat/0512696}
  (\bibinfo{year}{2005}).

\bibitem[{\citenamefont{Mukhamedshin et~al.}(2005)\citenamefont{Mukhamedshin,
  Alloul, Collin, and Blanchard}}]{Mukhamedshin:2005}
\bibinfo{author}{\bibfnamefont{I.~R.} \bibnamefont{Mukhamedshin}},
  \bibinfo{author}{\bibfnamefont{H.}~\bibnamefont{Alloul}},
  \bibinfo{author}{\bibfnamefont{G.}~\bibnamefont{Collin}}, \bibnamefont{and}
  \bibinfo{author}{\bibfnamefont{N.}~\bibnamefont{Blanchard}},
  \bibinfo{journal}{Phys. Rev. Lett.} \textbf{\bibinfo{volume}{94}},
  \bibinfo{pages}{247602} (\bibinfo{year}{2005}).

\bibitem[{\citenamefont{Zandbergen et~al.}(2004)\citenamefont{Zandbergen, Foo,
  Xu, Kumar, and Cava}}]{Zandbergen:2004}
\bibinfo{author}{\bibfnamefont{H.~W.} \bibnamefont{Zandbergen}},
  \bibinfo{author}{\bibfnamefont{M.}~\bibnamefont{Foo}},
  \bibinfo{author}{\bibfnamefont{Q.}~\bibnamefont{Xu}},
  \bibinfo{author}{\bibfnamefont{V.}~\bibnamefont{Kumar}}, \bibnamefont{and}
  \bibinfo{author}{\bibfnamefont{R.~J.} \bibnamefont{Cava}},
  \bibinfo{journal}{Phys. Rev. B} \textbf{\bibinfo{volume}{70}},
  \bibinfo{pages}{024101} (\bibinfo{year}{2004}).

\bibitem[{\citenamefont{Kresse and Furthmuller}(1996)}]{Kresse:1996}
\bibinfo{author}{\bibfnamefont{G.}~\bibnamefont{Kresse}} \bibnamefont{and}
  \bibinfo{author}{\bibfnamefont{J.}~\bibnamefont{Furthmuller}},
  \bibinfo{journal}{Phys. Rev. B} \textbf{\bibinfo{volume}{54}},
  \bibinfo{pages}{11169} (\bibinfo{year}{1996}).

\bibitem[{\citenamefont{Singh and Kasinathan}(2006)}]{Singh:2006}
\bibinfo{author}{\bibfnamefont{D.~J.} \bibnamefont{Singh}} \bibnamefont{and}
  \bibinfo{author}{\bibfnamefont{D.}~\bibnamefont{Kasinathan}},
  \bibinfo{journal}{Cond-mat/0604002}  (\bibinfo{year}{2006}).

\bibitem[{\citenamefont{Kotliar et~al.}(2006)\citenamefont{Kotliar, Savrasov,
  Haule, Oudovenko, Parcollet, and Marianetti}}]{Kotliar:2006}
\bibinfo{author}{\bibfnamefont{G.}~\bibnamefont{Kotliar}},
  \bibinfo{author}{\bibfnamefont{S.~Y.} \bibnamefont{Savrasov}},
  \bibinfo{author}{\bibfnamefont{K.}~\bibnamefont{Haule}},
  \bibinfo{author}{\bibfnamefont{V.~S.} \bibnamefont{Oudovenko}},
  \bibinfo{author}{\bibfnamefont{O.}~\bibnamefont{Parcollet}},
  \bibnamefont{and} \bibinfo{author}{\bibfnamefont{C.~A.}
  \bibnamefont{Marianetti}}, \bibinfo{journal}{Cond-mat/0511085}
  (\bibinfo{year}{2006}).

\bibitem[{\citenamefont{Georges et~al.}(1996)\citenamefont{Georges, Kotliar,
  Krauth, and Rozenberg}}]{Georges:1996}
\bibinfo{author}{\bibfnamefont{A.}~\bibnamefont{Georges}},
  \bibinfo{author}{\bibfnamefont{G.}~\bibnamefont{Kotliar}},
  \bibinfo{author}{\bibfnamefont{W.}~\bibnamefont{Krauth}}, \bibnamefont{and}
  \bibinfo{author}{\bibfnamefont{M.~J.} \bibnamefont{Rozenberg}},
  \bibinfo{journal}{Rev. Mod. Phys.} \textbf{\bibinfo{volume}{68}},
  \bibinfo{pages}{13} (\bibinfo{year}{1996}).

\bibitem[{\citenamefont{Parcollet and Marianetti}(2006)}]{Parcollet:2006}
\bibinfo{author}{\bibfnamefont{O.}~\bibnamefont{Parcollet}} \bibnamefont{and}
  \bibinfo{author}{\bibfnamefont{C.~A.} \bibnamefont{Marianetti}},
  \bibinfo{journal}{http://dmft.rutgers.edu}  (\bibinfo{year}{2006}).

\bibitem[{\citenamefont{Ishida et~al.}(2005)\citenamefont{Ishida, Johannes, and
  Liebsch}}]{Ishida:2005}
\bibinfo{author}{\bibfnamefont{H.}~\bibnamefont{Ishida}},
  \bibinfo{author}{\bibfnamefont{M.~D.} \bibnamefont{Johannes}},
  \bibnamefont{and} \bibinfo{author}{\bibfnamefont{A.}~\bibnamefont{Liebsch}},
  \bibinfo{journal}{Phys. Rev. Lett.} \textbf{\bibinfo{volume}{94}},
  \bibinfo{pages}{196401} (\bibinfo{year}{2005}).

\end{thebibliography}

\end{document}